\documentclass[iicol,sn-basic]{sn-jnl}
\usepackage{float}
\usepackage[section]{placeins}
\usepackage[finalnew]{trackchanges}

\jyear{2023}

\theoremstyle{thmstyleone}

\theoremstyle{thmstyletwo}

\theoremstyle{thmstylethree}

\begin{document}

\title{High-precision interferometric measurement of \change{rapid}{slow and fast} temperature changes in \add{static fluid and} convective flow}

\author*[1]{\fnm{Xinyang} \sur{Ge}}\email{xg2n17@soton.ac.uk}

\author[1,2]{\fnm{Joanna A.} \sur{Zieli\'{n}ska}}\email{jzielinska@ethz.ch}

\author*[1]{\fnm{Sergio} \sur{Maldonado}}\email{s.maldonado@soton.ac.uk}

\affil[1]{\orgdiv{Faculty of Engineering and Physical Sciences}, \orgname{University of Southampton}, \orgaddress{\city{Southampton}, \postcode{SO16 7QF}, \country{UK}}}

\affil[2]{\orgdiv{Photonics Laboratory}, \orgname{ETH Zurich}, \orgaddress{\city{Zurich}, \postcode{CH-8093}, \country{Switzerland}}}

\abstract{\add{We explore the strengths and limitations of using a standard Michelson interferometer to sample line-of-sight-averaged temperature in water via two experimental setups: slow-varying temperature in static fluid and fast temperature variations in convective flow. } \remove{Using interferometry, we measure rapid temperature variations in convective flow to within a few mK. This} \add{The high precision of our measurements (a few mK)} is enabled by the fast response time and high sensitivity of the interferometer to minute changes in the refractive index of \change{the test fluid}{water} caused by temperature variations. These features allow us to detect the signature of fine fluid dynamical \change{patters}{patterns} in convective flow in a fully non-intrusive manner. For example, we are able to observe an asymmetry in the rising thermal plume (i.e. an asynchronous arrival of two counter-rotating vortices at the measurement location), which is not possible to resolve with more traditional (and invasive) techniques, such as RTD (Resistance Temperature Detector) sensors. These findings, and the overall reliability of our method, are further corroborated by means of Particle Image Velocimetry and Large Eddy Simulations. \change{The non-intrusive nature of this method, along with its robustness and}{While this method presents inherent limitations (mainly stemming from the line-of-sight-averaged nature of its results), its non-intrusiveness and robustness, along with} the ability to readily yield real-time, highly accurate measurements, render this technique very attractive for a wide range of applications in experimental fluid dynamics.}

\keywords{Interferometry, Convective flow, Optical metrology, Temperature measurement}

\maketitle

\section{Introduction}\label{secIntro}

Reliable measurement of temperature in fluids is important for a wide range of applications, from industrial (e.g. power plants, heat exchangers, chemical industries) to basic scientific research \citep[e.g.][]{ross2001temperature,abram2018temperature,laffont2018temperature}. Fluid temperature is conventionally measured via intrusive instrumentation. Thermocouples, thermistors and resistance temperature detectors (RTDs) are popular examples of instruments which measure fluid temperature by immersion of a physical probe. While this may not be problematic for many industrial applications, when studying fluid dynamics from a scientific perspective, this intrusiveness may be undesirable since the probes may disturb local flow patterns. Also, these sensors typically have moderate to poor accuracy. While RTDs are generally considered accurate for most applications, their slow response time prevents them from capturing fast changes in fluid temperature \citep{Goumopoulos}, of the type expected in turbulent flows. Another class of temperature sensors are based on optical fibers \citep{childs2000review,Fernandez-Valdivielso,Rizzolo,Kim2018}, which offer advantages such as small dimensions, capability of multiplexing, chemical inertness, and immunity to electromagnetic fields \citep{roriz2020optical}. It must be noted, however, that while optical fibre based sensors can be as small as a fraction of a millimetre in diameter \citep{drusova2021comparison}, they must be immersed in the test fluid, making them invasive instruments. Another intrusive technique is acoustic thermometry, which relies on the measurement of the speed of sound in a given fluid \citep{moldover2014acoustic,wang2018acoustic}. The response time of this technique is determined by the time traveled by a probe sound wave, in turn given by the thickness of the sample volume. This method can attain a precision of up to $\pm$0.015 K with a sampling frequency of 5 Hz \citep{wang2018acoustic}.

When it is not possible or desirable for sensors to be in direct contact with the test fluid, a non-intrusive method may be used, such as laser-induced fluorescence (LIF) or interferometry. 
LIF (originally proposed by \citealt{tango1968spectroscopy}) works by using laser light to excite a fluorescent dye, which subsequently fluoresces at a different wavelength, revealing temperature changes \citep{banks2019planar}. \add{LIF has been employed in numerous studies to understand heat transfer in convective flow }\citep[e.g.][]{grafsronningen2012simultaneous,Park,kashanj2023application}, and can be used to obtain both two- and three-dimensional images \citep{taylor2021current}, with accuracy in thermofluid applications of up to 0.17 K \citep{sakakibara2004measurement}. 

Interferometric techniques for visualisation and analysis of transparent fluid flows are non-invasive, fast and extremely sensitive. These include holographic interferometry (HI) and standard interferometry (SI). Inteferometric methods extract fluid properties (e.g. temperature) from spatial and/or temporal interference between two laser beams, one of which passes through the test fluid (the other one being the reference). In HI, the interference pattern between two laser beams is recorded on a camera and used to reconstruct the optical phase distribution in the observed area, which directly maps onto \add{the} 2D temperature distribution within the field of view of the instrument. 
Several authors have employed HI to understand convection-based heat transfer phenomena around horizontally placed heating rods \citep{herraez2002study,ashjaee2007experimental1,ashjaee2007experimental,narayan2017interferometric} as well as temperature profiles during liquid cooling and heating processes \citep{wu2013visual,guerrero2016real,wang2022single}. These investigations have successfully extracted temperature variations in the fluid, with reported accuracies ranging from 0.20 K \citep{guerrero2016real} to 0.42 K \citep{wu2013visual}. 
Most studies have been limited to laminar flows, and the reported experimental data are recorded under steady-state conditions after the initial transients have \change{died out}{decayed}. This may be due to the fact that, in dynamic measurements, HI performance is limited, as large phase gradients across the beam (which may occur in turbulent flow settings) may lead to non-distinguishable interference structure and errors in the interpretation of fringe patterns~\citep{Wilkie63, Lira87}. 

Unlike HI, SI does not provide spatial information about the distribution of fluid properties such as temperature. Instead, in SI the temporal interference between probe and reference beams is observed using a single-pixel detector, which provides necessary information to measure temporal variations of the optical path length in one location. Continuous modulation of the reference arm length enables monitoring of the interference visibility, which provides information about the performance of the instrument (which may be affected by phase gradients perpendicular to the probe beam's direction of propagation). This is advantageous for highly non-uniform flow settings, permitting measurements in the presence of large phase gradients (albeit with reduced sensitivity). 
What is more, SI is relatively simple and economic, with data processing requirements being significantly less computationally demanding than those of HI. These characteristics are desirable for real-time measurements of rapid temperature variations in fluids. 

\add{While SI is a well-established technique with a wide range of applications in metrology and other branches of science} \citep[e.g.][]{smith1989absolute,freise2009triple,ito2020interferometric}, \add{to the best of the authors' knowledge, very few researchers have employed this technique to sample temperature in fluid dynamical setups. For example,} \cite{Tomita} and \cite{mahdieh2013measurement} \add{implemented SI to measure very slow water temperature changes in uniformly heated water. However, their detection method (involving the imaging of interference fringes using a camera) enables only a modest measurement accuracy; e.g. of $\sim$6\% in the case of} \cite{mahdieh2013measurement}. \cite{zhang2019water} \add{used SI to measure salinity and temperature in slowly cooling water (over a 10 hour period). Their method, necessitating empirical expressions (relating temperature, salinity and refractive index), yielded measurements with an accuracy of 0.12 K relative to the reference instrument (a thermistor).}

In this paper, we \add{explore the potential and limitations of SI in experimental fluid dynamics by applying it to the study of two different problems: slow-varying temperature in static water and convective flow in water induced by a horizontal heating rod. The latter, in particular, is a novel  application of SI. Results are compared against and complemented by well-established experimental and numerical methods, such as RTD sensors, Particle Image Velocimetry and Large Eddy Simulations.} \remove{apply SI as a non-intrusive, fast-response method to study convective flow in water induced by a horizontal heating rod. We are able to characterise the transient heat transfer phenomenon, including the raising thermal plume (and consequent large temperature gradients), and observe the signature of fluid dynamical patterns which are too fast to be recorded by other methods such as RTD sensors.} The structure of the paper is as follows. 
In Section \ref{secMethod}, we explain the experimental setup and the procedure to extract temperature changes from shifts in interference patterns. 
Then, Section \ref{secStatic} presents the results of a slowly varying temperature experiment in \add{a} static fluid, which establishes the method's reliability and sensitivity. 
Next, in Section \ref{Convective flow}, we apply our technique to \add{a} convective flow, and compare the obtained results against those from RTD sensors, further enhancing our analysis by \add{the} use of Particle Image Velocimetry and Large Eddy Simulations. 
\add{Section} \ref{secDiscussion} \add{presents a brief discussion on the key merits and limitations of the proposed technique, as well as an outlook for potential future applications and improvements.}
Concluding remarks are presented in Section \ref{secConclusions}.

\section{Methodology}\label{secMethod}

\subsection{Interferometry: basic theory}

\begin{figure*}[t]
    \centering
    \includegraphics[width=0.95 \textwidth]{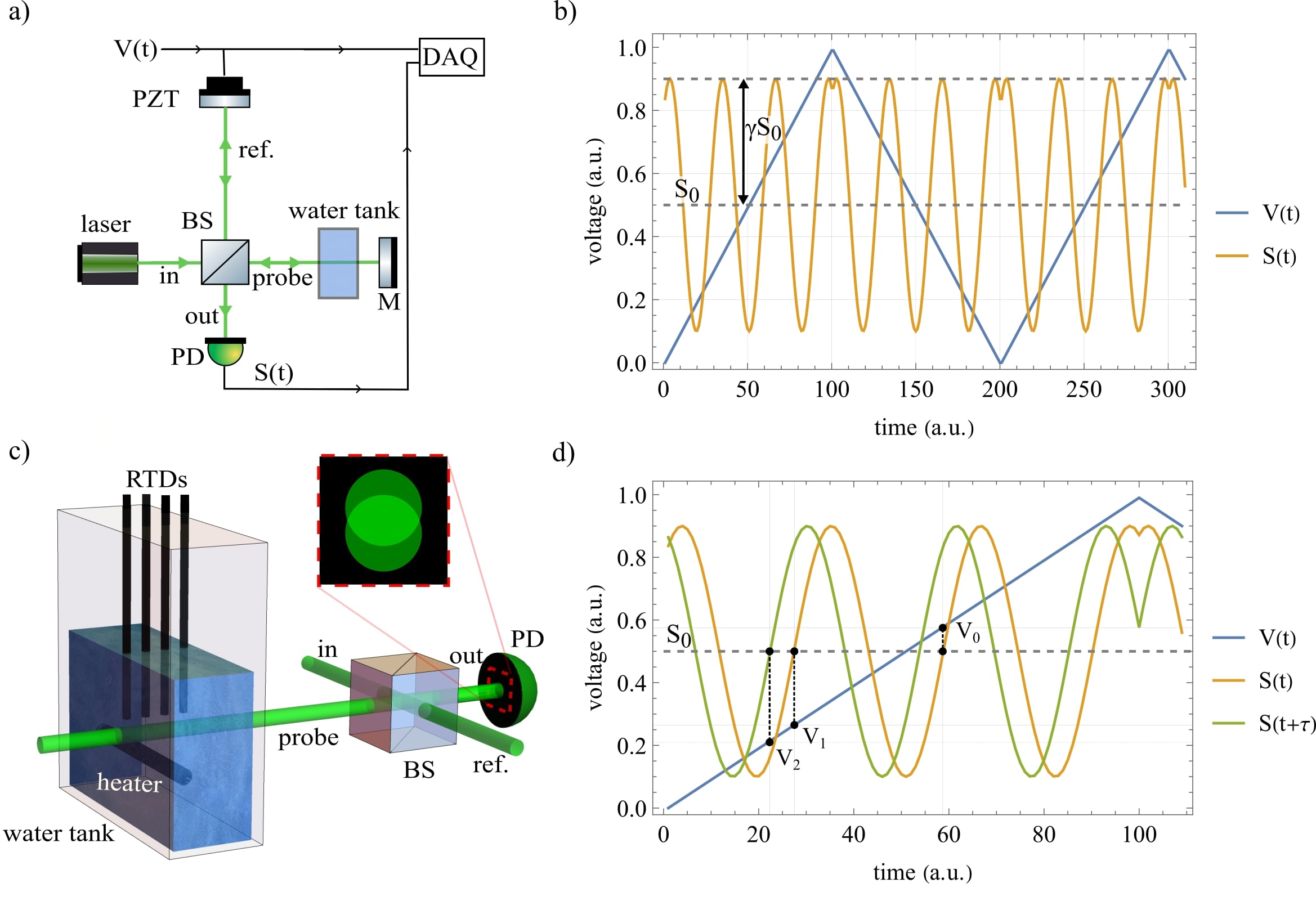}
    \caption{{\bf a)} Experimental setup of the interferometer. The reference arm length is controlled using a piezoeletric actuator driven by a voltage signal $V(t)$. The probe arm passes twice through the water tank.  Abbreviations: PZT - piezoeletric actuator, M - mirror, BS - non-polarising 50:50 beamsplitter, PD - photodetector, DAQ - data acquisition device. {\bf b)} Oscilloscope view of the PZT driving voltage, $V(t)$, and signal recorded on the detector PD, $S(t)$. {\bf c)} 3D render of the setup, showing approximate position of the heating rod and RTD sensors inside the water tank, as well as a close-up schematic view of the overlap between probe and reference beams at the PD. {\bf d)} Illustration of the method for extracting temperature changes from interference pattern shifts; $S(t)$ and $S(t+\tau)$ are PD-detected signals during subsequent piezo strokes.}
    \label{fig:simple_setup}
\end{figure*}

In most interferometers, light from a single coherent source is split into two beams that travel via different optical paths, which are then recombined to produce an interference pattern. The resulting interference fringes provide information about the difference in optical path lengths with a precision of a fraction of the wavelength of light.

Here, we measure water temperature using a Michelson interferometer. The probe arm of the interferometer passes through the water tank, while the second (reference) arm path bypasses the tank. The interferometer measures the phase difference between the two arms, which depends on the refractive index of water, in turn given by its temperature. 

The experimental setup is shown in Fig.~\ref{fig:simple_setup}(a). A laser beam is split into two equal-power components with the help of a beam-splitter. One of the beams passes through the water tank and is reflected off a stationary mirror, while the other one is reflected off a mirror mounted on a piezoelectric actuator, fed by a voltage signal shown in Fig.~\ref{fig:simple_setup}(b). The interference signal $S(t)$ is detected by a photodetector. The signal $S(t)$ consists of fringes, resulting from linear phase ramps (i.e. linear piezo displacement in time; see $V(t)$ curve in Fig.~\ref{fig:simple_setup}b) introduced by the piezoelectric transducer (PZT). Each time the PZT reverses direction, we observe a `turning point' in the fringe pattern (see times 100, 200 and 300 in Fig.~\ref{fig:simple_setup}b). \add{It must be noted that the probe beam passes twice through the test fluid, which effectively increases the instrument's sensitivity by a factor of $2$ (compared to a single-pass configuration).}

Interferometry does not provide absolute temperature measurements, but rather variations with respect to time. We detect time variations of the temperature (refractive index) of water by comparing the fringe patterns recorded in different moments in time (consecutive PZT sweeps). A shift of the fringe pattern with respect to the PZT drive signal allows us to deduce the phase difference between the arms, $\phi(t)$, which is expressed as: 

\begin{equation}
    \phi(t)= k_0 z_{\rm pzt}(t)- 2 k_0 \int_0^l n_{\rm w} (t) \,dz +\phi_0(t),
    \label{eq:phase}
\end{equation}
where  $k_0$ is the wave-vector in vacuum, $k_0 z_{\rm pzt}(t)$ is the phase shift introduced by the displacement of the PZT (proportional to $V(t)$), the $z$ coordinate is aligned with the direction of beam propagation through the water tank along a path of length $l$, and $n_{\rm w} (t)$ is the refractive index of water (which may vary in time). The term $\phi_0$ contains mechanical drifts which affect the lengths of the interferometer arms, as well as phase difference introduced by variations in air temperature. 

For small temperature variations $\delta T = T_1 - T_0$, we can approximate the dependence of water refractive index on temperature $T_1$ as

\begin{equation}
    n_{\rm w}(T_1)\approx n_{\rm w}(T_0) + \left. \left(\frac{\partial n}{\partial T}\right) \right\rvert_{T_0} \delta T\, ,
    \label{eq:refindex}
\end{equation}

where $\partial n_{\rm w} / \partial T = -9.6\times 10^{-5}~\rm{K}^{-1}$ within the range of 285 to 305 K at 101,325 Pa  \citep{fernandez1997release}. Note that $\partial n / \partial T$ for air is approximately $-9\times 10^{-7}~{\rm K}^{-1}$ \citep{Ciddor:96} (dry air, 450 ppm CO$_2$), such that the experiments described in this paper are much more sensitive to water than to air temperature changes (by two orders of magnitude) that may occur in the lab. \add{Indeed, this is why we select water as the test fluid.}

Note that the temperature variation deduced from $\phi(t)$ corresponds to the average change along the beam path (inside the water).

The signal $S(t)$ detected by the photodiode (see PD in Fig.~\ref{fig:simple_setup}) and $\phi(t)$ are related via:

\begin{equation}
    S(t)= \eta \left[ I_1+I_2 + 2 \sqrt{I_1 I_2} \gamma  \cos(\phi(t))\right],    
\end{equation}
where $I_1$ and $I_2$ are the optical intensities corresponding to the probe and reference beams, respectively, and $\gamma$ is the absolute value of the first order normalized correlation function between probe and reference fields~\citep{mandel_wolf_1995}. The calibration factor $\eta$ converts the signal to Volts.  
The two beams have approximately equal power, such that: 

\begin{equation}
     I_1 \approx I_2 = \frac{S_0}{2\eta}\, ,
\end{equation}
where $S_0$ is the DC part of the signal $S(t)$ (see Fig. \ref{fig:simple_setup}b). The amplitude of the interference fringes relative to $S_0$ (interference visibility) is dictated by $\gamma$. Unit visibility ($\gamma=1$) occurs when the spatial coherence (overlap between probe and reference beams) is perfect and the path length difference between probe and reference beams is much smaller than the coherence length. In this work the latter is always true, i.e. our anticipated optical path length differences are always well within the coherence length, which for our laser source is equal to the path length difference resulting from a water temperature change of 59 K. The interference visibility in our experiment is therefore limited by spatial misalignment between the two beams and wavefront distortion introduced by water, which results in imperfect overlap of probe and reference beams at the detector (see inset in Fig.~\ref{fig:simple_setup}c).

The method for extracting temperature changes from the fringe pattern is depicted in Fig.~\ref{fig:simple_setup}d. We calibrate the PZT using the fact that a change of (double-pass) path length difference of one wavelength is equivalent to the shift of the sinusoidal signal pattern by one period. Therefore, a voltage change of $\vert V_1-V_0\vert$ (see Fig.~\ref{fig:simple_setup}d) corresponds to a single-pass path length difference of $\lambda/2$, yielding a calibration factor $\alpha = \frac{\lambda}{2 \vert V_1-V_0\vert}$. The temporal change in the refractive index (integrated over $l$) corresponding to a voltage change $V_2 - V_1$ can then be obtained from: 

\begin{equation}
\int_0^l \left( n_{\rm w} (t+\tau)- n_{\rm w}(t) \right)\,dz = \alpha (V_2 - V_1) ,
\end{equation}
where $V_1$ is the `$S_0$-crossing' of the signal at time $t$ and $V_2$ corresponds to the subsequent piezo stroke at time $t+\tau$ (see Fig.~\ref{fig:simple_setup}d).
Finally, we can calculate the temperature change as:

\begin{equation}
\widehat{\delta T} = \widehat{T}(t+\tau)- \widehat{T}(t) = \frac{\alpha}{l} (V_2 - V_1) \left( \frac{\partial n_w}{\partial T} \right)^{-1} ,
\end{equation}
where $\widehat{T}(t)$ denotes average water temperature along the beam \add{(corresponding to the average refractive index along the beam) and $\widehat{\delta T}$ is the corresponding line-of-sight-average change in temperature (but, for clarity, we drop the hat notation hereinafter)}. In our experiments, we sample temperature changes sufficiently fast so that $ \vert V_2-V_1\vert < \vert V_1-V_0\vert $ is always satisfied.

\section{Static fluid experiment } \label{secStatic}
As a first assessment of the proposed technique, we compare its performance against that of state-of-the-art RTD (Resistance Temperature Detector) sensors for a long experiment where water temperature varies slowly. In order to quantify the stability and sensitivity of \change{our}{the} technique, we also analyse its Allan variance.

\subsection{Experimental setup}
A schematic of the experimental setup is illustrated in Fig. \ref{fig:simple_setup}. The outer dimensions of the glass tank are $0.20 \times 0.10 \times 0.30$ m in length, width and height, respectively. We use a collimated laser with a wavelength of 532 nm \add{(Thorlabs, CPS532)} and orient the tank such that the laser travels along its length. \add{The beam diameter is approximately 3 mm.} The beam splitter is a non-polarising cube beam splitter with a 50:50 (Reflectance : Transmission)  beam splitting ratio and the moveable mirror is powered by a discrete piezo ring stack \add{(Thorlabs, PK44LA2P2)} with a maximum displacement of 9 $\mu$m when driven with no load. The piezo actuator is powered by a waveform generator \add{(ISO-TECH, GFG-8216A)}, which can produce different types of waveforms (e.g. triangular, sinusoidal) with varying frequencies. Four Class 1/10 DIN RTD sensors \add{(OMEGA, P-M-1/10-1/4-6-0-P-3)} are evenly spaced inside the water with their probes located near the laser path for comparison against the \add{line-of-sight-averaged} results obtained from the interferometer; \add{for the same reason, all RTD measurements reported in this paper refer to the average of all four sensors}. The RTDs are connected to a data logger \add{(DataTaker DT85M)} for recording temperature with a 4-wire configuration. Their accuracy is $\pm (0.03+0.0005 T_s) \, { }^ \circ C$, where $T_s$ is the surrounding fluid temperature.

The tank is filled with 20 cm of water and the laser beam \change{is placed}{propagates horizontally at} 5 cm from the bottom of the tank. For this experiment, the piezo is driven with a triangular signal at 40 Hz. The experiment runs for 13 hrs in order to compare the results from the proposed technique against measurements from RTD sensors in a simple setting where water temperature varies slowly due to ambient temperature changes. The experiments have been performed remotely at night hours to minimise mechanical vibrations induced by human movement. A \change{layer of insulation}{vibration absorption mat} is also placed underneath the optical table to further reduce possible noise.

\subsection{Results} \label{secConvResults}

\begin{figure}[ht]
  \centering
  \includegraphics[width=1.0 \linewidth]{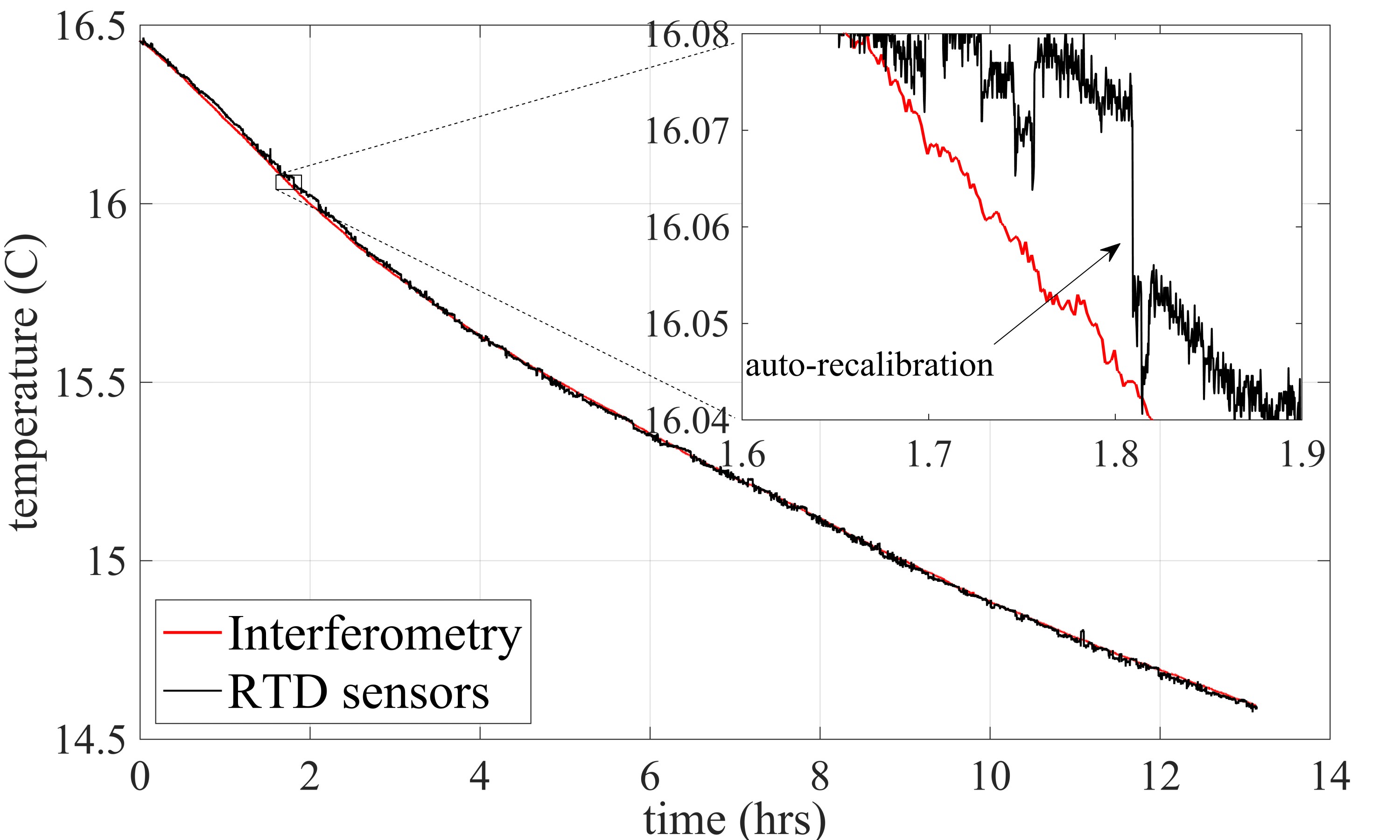}
  \caption{Comparison of measurements from RTD sensors and interferometry for slowly varying water temperature in static fluid. Differences are barely noticeable (see inset).}
  \label{fig:comparison}
\end{figure}

The temperature variations measured by the RTD sensors and inferred from interferometry are compared in Fig. \ref{fig:comparison}. The observed drop in water temperature is entirely due to the drop in room temperature over night (no additional cooling system is employed). The sampling frequency of the data logger for the RTD sensors is 1 Hz, and the reported value is the average of the four sensors. For the interferometry results, we captured a triggered result based on the voltage from the waveform generator every 6 seconds. The data transmission rate of the oscilloscope \add{(Tektronix, TDS-2004C)} limits this interval. While a different data logger could have been employed for continuous data recording (see Section \ref{set-up}), the volume of data corresponding to this 13 hrs long experiment would have been unnecessarily large for post-processing purposes. Recalling that the proposed technique yields temperature changes rather than absolute values, we have fixed the reference temperature at $t=0$ as equal to that from the RTD sensors \add{(i.e. their average)}. This is the only calibration we carry out. 
Fig. \ref{fig:comparison} shows that the two curves agree extremely well; the difference between the methods has an average of only $2.47 \times 10^{-4}$ K. Most noticeable differences occur right before the auto-recalibration process of the data logger (see inset in Fig. \ref{fig:comparison}). During operation, the RTD data logger measures the amplifier's internal `offset voltage'. If it is found to have drifted by a specified amount ($1\mu V$), a calibration cycle is performed \citep{dataTaker}. Regardless of the error caused by the data logger drift, the difference between the two methods is within the expected accuracy of the RTD sensors ($\sim 0.038$ K). This comparison against state-of-the-art temperature sensors in a simple setting serves as a confirmation of the high accuracy achieved by the proposed method.

\subsection{Interferometer stability} \label{allan 1}

We quantify the stability of the interferometer under hydrostatic conditions, which determines the \change{sensitivity}{precision} of our measurements for a given integration time. A standard method to characterise a sensor's stability is the two-sample variance, also known as Allan variance~\citep{allan}, which quantifies noise in the system and resulting measurement limitations. For the derivation of the two-sample variance $\sigma^2_T(\tau)$ of a time trace $T(t)$ with length $\mathcal{T} $, the time trace is split into $N$ equal blocks of length $\tau= \mathcal{T} /N$. For each of these blocks an average value is denoted as $\bar{T}_k^{(\tau)}$, where $k\in (1,...,N)$. The Allan variance is the expectation value of the two consecutive values of the block average $\bar{T}_{k}^{(\tau)}$ as a function of integration time $\tau$:

\begin{align}
\sigma^2_T(\tau)&=\left<\frac{1}{2}\left(\bar{T}_{k+1}^{(\tau)}-\bar{T}_{k}^{(\tau)}\right)^2\right>\nonumber\\
&\approx \frac{1}{2(N-1)}\sum_{k=1}^{N-1}\left(\bar{T}_{k+1}^{(\tau)}-\bar{T}_{k}^{(\tau)}\right)^2 .
\end{align}

 \begin{figure}[ht]
  \centering
  \includegraphics[width= 1.0 \linewidth]{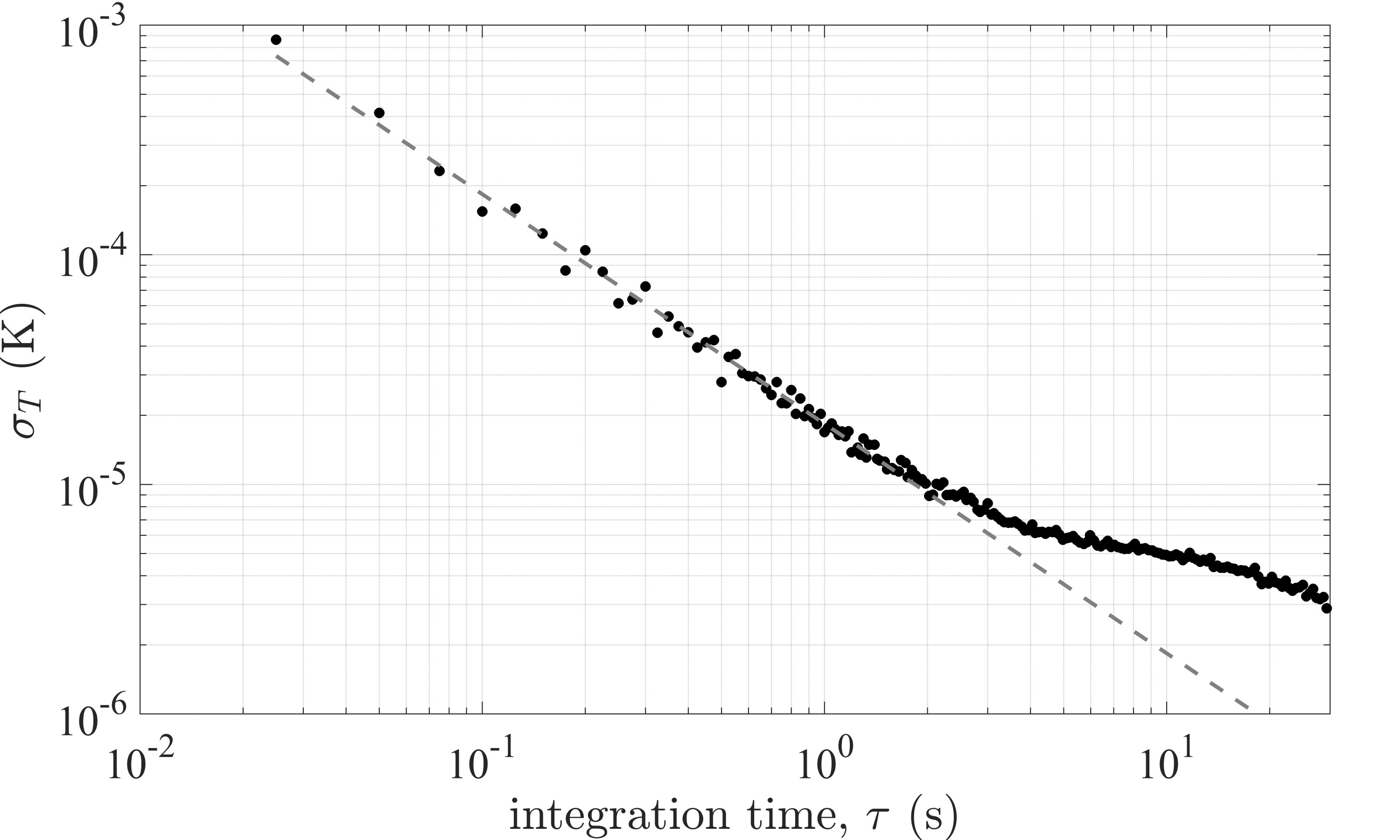}
  \caption{Allan deviation plot for the static fluid setting; slope $\tau^{-1}$ shown as dashed line.  }
  \label{fig:allan}
\end{figure}

The result of the Allan variance analysis is shown in Fig.~\ref{fig:allan}, which has been calculated by acquiring data for \change{2 hrs}{1 min} with a $40$~Hz sampling frequency. We observe that for integration times shorter than approximately 2 s, the signal behaviour is dominated by high frequency noise. Increasing the averaging window beyond $\sim 2$ s reveals the influence of other noise sources, including long-term mechanical drifts in the optical path lengths. We estimate that the \change{sensitivity}{precision} of our instrument is \change{of}{approximately} $1$~mK per piezo cycle (at 40 Hz), and up to $5~{ \rm\mu K}$ for 10 s integration time. \add{Under the assumption that the water temperature remains constant over a $1/40$ s interval (a reasonable assumption given the experimental conditions), $1$~mK can also be taken as a good estimate of the accuracy of the interferometer (at a 40 Hz sampling rate).}
 
\section{Convective flow experiment}
\label{Convective flow}

In this section we describe the measurements of rapid temperature variations due to natural convection induced by a heating rod in the fluid. Interferometry results are compared against those from RTD sensors. Our interpretation of the measurements obtained is enhanced via numerical simulations and Particle Image Velocimetry.

\subsection{Experimental setup}
\label{set-up}

\begin{figure}[ht]
  \centering
  \includegraphics[width=6cm]{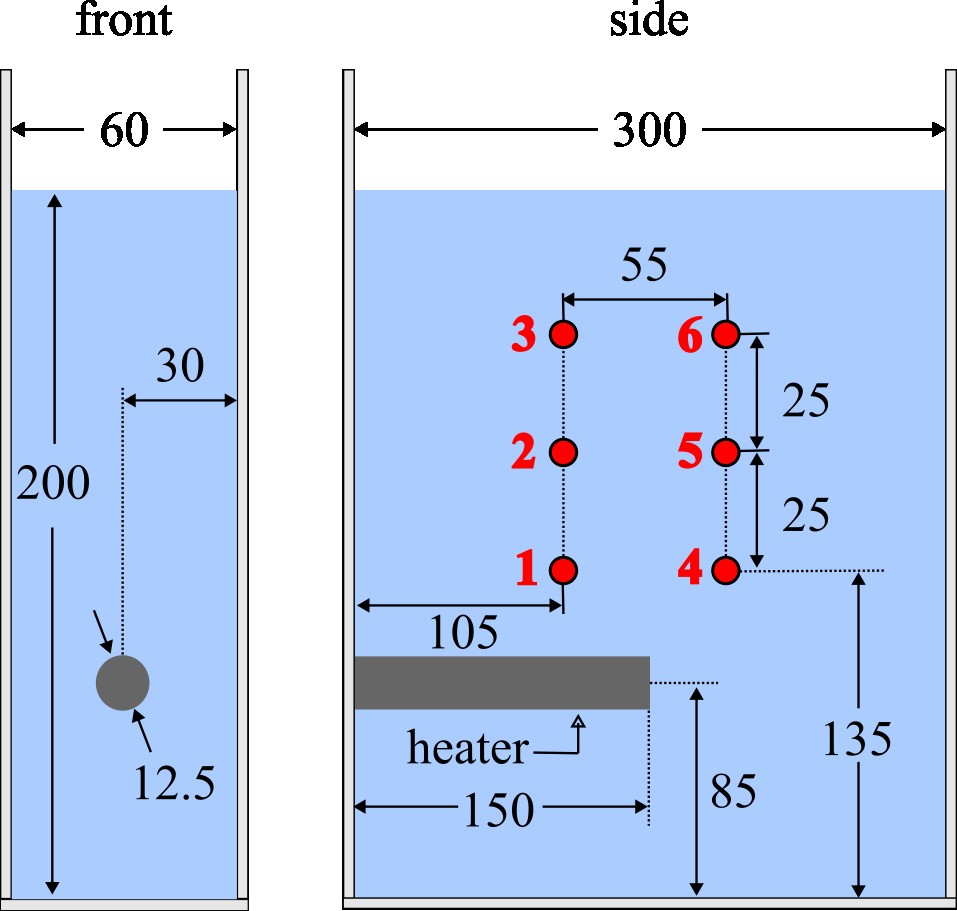}
  \caption{Sketch of the water tank used in the convective flow experiments showing geometrical details and location of all six \change{points}{locations} sampled (in red). All dimensions in mm. }
  \label{fig:points}
\end{figure}

The experimental setup is depicted in Figure \ref{fig:simple_setup}. The data acquisition device used for this experiment is a Red Pitaya 125-10 (a 10-bit data logger), which facilitates continuous high-frequency data streaming for rapidly fluctuating water temperature measurements. The tank is positioned on a jack lift, with a rectangular opening in the optical table allowing for horizontal and vertical adjustments to sample data from various locations without necessitating laser realignment. 
The tank geometry is based on that reported by \cite{Park} and is shown in Fig. \ref{fig:points}. The material is common glass with a nominal thickness of 10 mm. The tank is filled with water up to 20 cm and the power output from the heater is set to 12 W. 
For analysis and comparison against RTD measurements, six different locations were chosen as illustrated in Fig. \ref{fig:points}. Some of these points are right above the heater, while others are slightly off. This is done to capture qualitatively different parts of the experiment; e.g. unlike locations 4-6, the area right above the heater (locations 1-3) is expected to be strongly dominated by the rising plume (see Fig.~\ref{fig:LES_IR}). 

As we show later, a short-duration experiment is not appropriate to compare interferometric measurements against those from RTD sensors due to the latter's slow response time. For this reason, experiments were conducted over a 480 s time span. Data collection began right after switching on the heater, which was then switched off at $t=80$ s, and data recording continued for another 400 s. Therefore, while instantaneous temperature is not expected to be the same for both methods (given their different response times), the final recorded temperature should in theory match (by design, we take the initial temperature to be the same in both methods).

\subsection{Results}

\begin{figure*}[ht]
  \centering
  \includegraphics[width= 0.97 \linewidth]{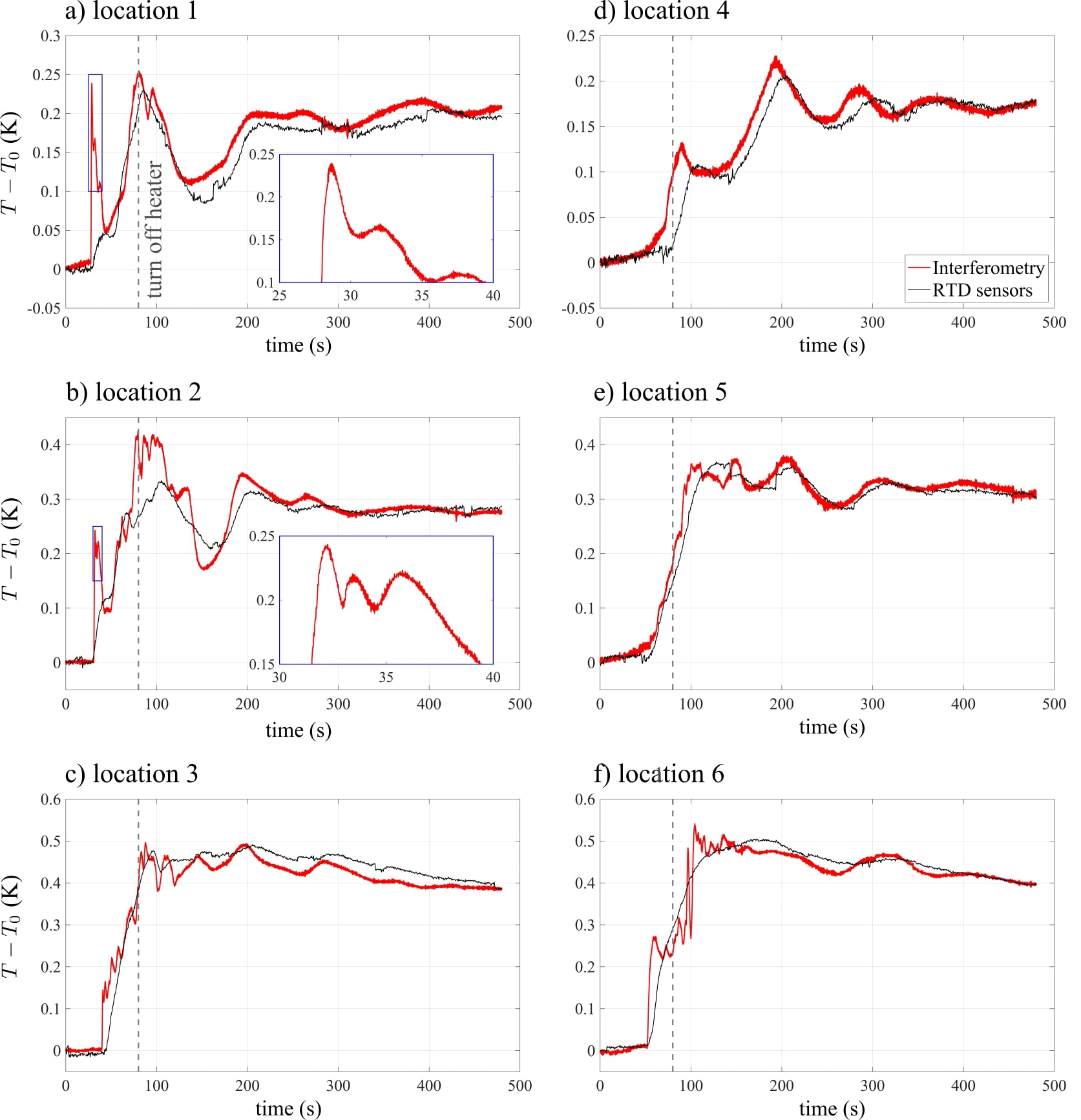}
  \caption{\add{Line-of-sight-average} temperature change relative to initial temperature, $T-T_0$, for the convective flow experiment. Vertical dashed line represents the time at which the heater was turned off. RTD data (black line) denotes the average of all 4 RTD sensors. Insets in a) and b) are further analysed in Section \ref{sec:PIV_LES} and Fig. \ref{simulation}.}
  \label{dynamic results}
\end{figure*}

\subsubsection{Comparison against RTD sensors} \label{sec:convec_vs_RTDs}
\begin{figure}[ht]
  \centering
  \includegraphics[width= 0.95 \linewidth]{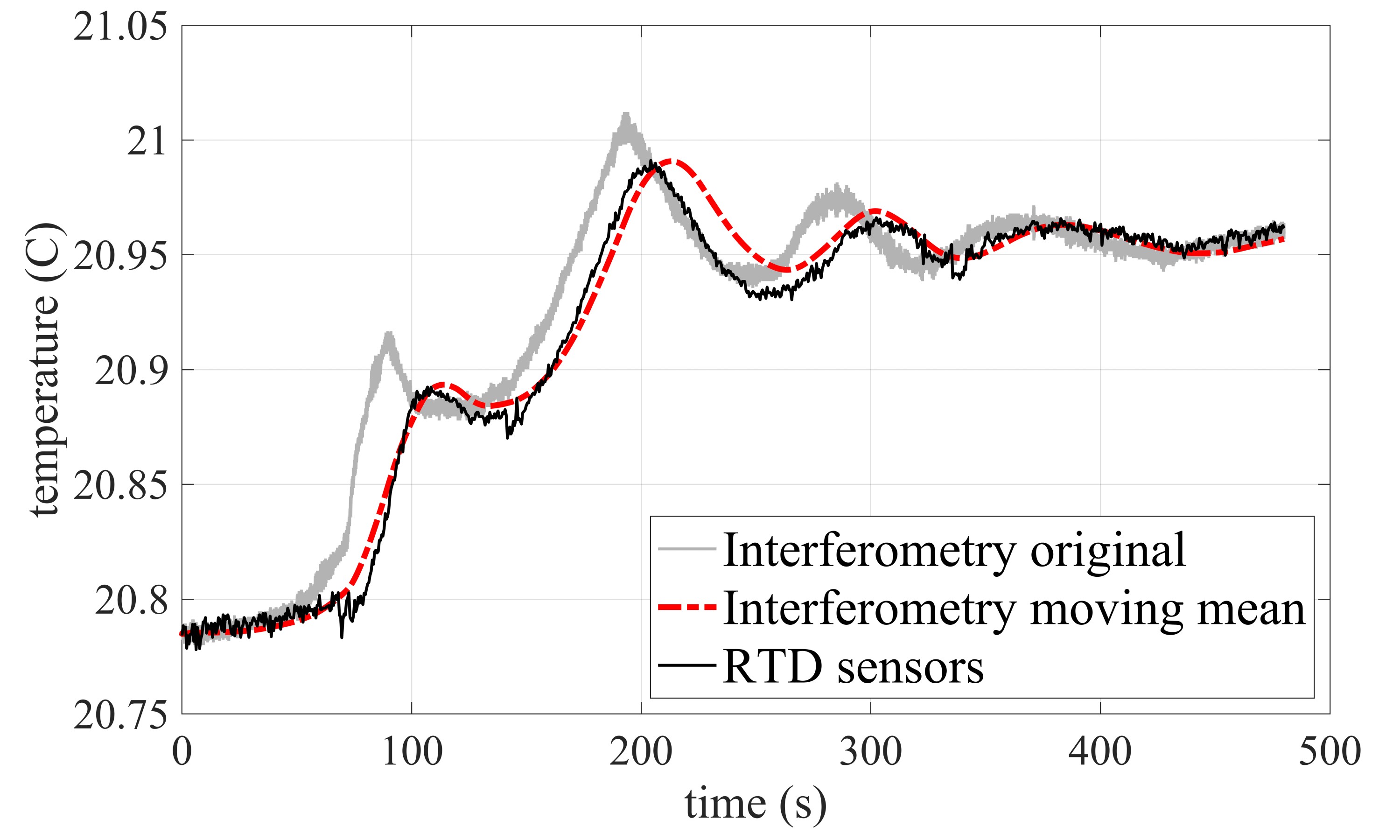}
  \caption{Measurements from RTDs and interferometry in location 4 after applying a simple moving average function to the interferometry results. Here, the moving-mean temperature reported (red dashed line) at time $t$ represents the average of the previous $30$ s.}
  \label{moving average}
\end{figure}

Comparison between the RTDs and interferometry in six locations within the tank for the convective flow experiment is illustrated in Fig. \ref{dynamic results}. 
Overall, both methods agree well (especially if the RTDs accuracy is taken into account, which is of $\pm$0.04 K). However, it is worth analysing several aspects in detail. First of all, all six sampled locations exhibit a high degree of agreement between both techniques at the end of the experiment. This is important because the proposed interferometric method works (i.e. yields instantaneous temperature) by accumulating temperature changes relative to the previous instant in time (piezo stroke), such that if major errors occurred, these would be evident from the final temperature measurement. We are thus confident that such error accumulation does not take place. Secondly, in all locations a lag between both methods is evident, with temperature peaks being smoothed out and recorded later by the RTDs (this is particularly clear in location 4). This is an expected behaviour given the well-known slow response time of RTDs. We can further verify this by applying a simple moving average function to the original interferometric measurements \add{(in this case, we report at time $t$ the average temperature over the previous $30$ s)}, which closely replicates the RTD data, as illustrated in Fig. \ref{moving average} (evidently, this agreement could improve if a different moving average filter, better mimicking the RTDs behaviour, were employed). 
Finally, notice that rapid temperature variations are not captured by the RTDs, which is especially evident in location 1, where interferometry reports a sharp rise of about 0.23 K within 1.0 s, followed by a decrease in temperature. This phenomenon, also observed in location 2, is presumably caused by the rising plume, and we further analyse it in Section \ref{sec:PIV_LES}. Temperature fluctuations are also present in locations 3 and 6, which we attribute to the plume's transition to turbulent flow as it approaches the free surface. Locations 4 and 5 do not exhibit the aforementioned discrepancy between techniques due to the the primary mode of heat transfer in these locations, which is conduction rather than convection, leading to the absence of the sharp temperature increases observed in other locations.

\subsubsection{Comparison against PIV and LES} \label{sec:PIV_LES}
The temperature variations measured in locations 1 and 2 at around $t=30$ s are particularly intriguing, as interferometry not only reveals a rapid increase in temperature due to the rising plume, but also uncovers three peaks within it (see insets in Figs. \ref{dynamic results}a and \ref{dynamic results}b). To investigate further the origin of these peaks, we analyse the setup using Particle Image Velocimetry (PIV) and Large Eddy Simulations (LES), focusing our attention on a cross-section normal to the heating rod passing through locations (beam paths) 1, 2 and 3, as illustrated in Fig. \ref{fig:LES_IR} (see Appendices \ref{piv appendix} and \ref{les appendix} for technical details on the PIV and LES settings, respectively).

 \begin{figure}[ht]
  \centering
  \includegraphics[width= 0.85 \linewidth]{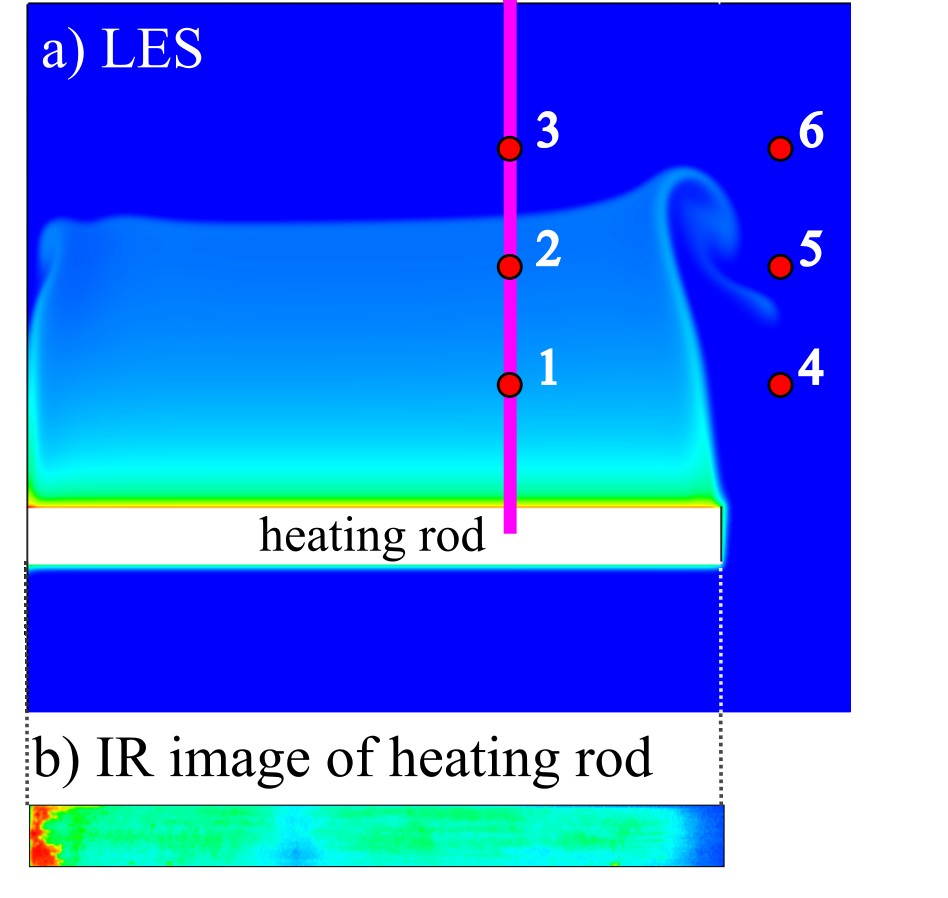}
  \caption{a) Illustration of a Large Eddy Simulation (temperature field), indicating the location of the cross-section of interest (purple vertical line), which has also been analysed by means of Particle Image Velocimetry (see Fig. \ref{simulation}). b) Non-uniform temperature distribution observed in the heating rod using an infrared (IR) camera. Images provided for qualitative analysis only (hence, no colour bar).}
  \label{fig:LES_IR}
\end{figure}

\begin{figure*}

  \centering
  \includegraphics[width= 0.86 \linewidth]{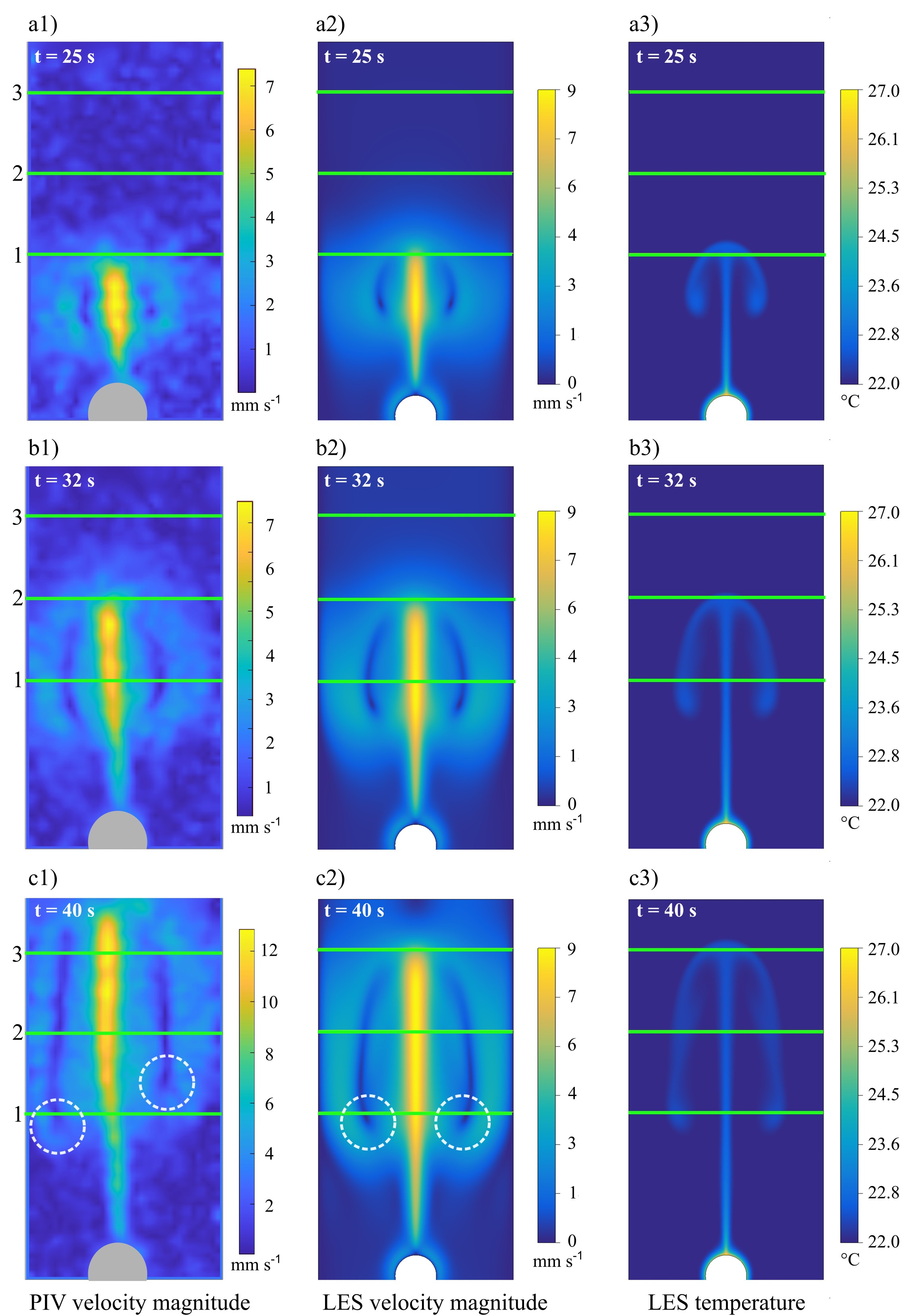}
  \caption{Velocity magnitude obtained via Particle Image Velocimetry (left panels) and Large Eddy Simulations (centre panels), as well as temperature contour plots from LES (right panels), at three different times after start of the experiment: a) 25 s, b) 32 s, and c) 40 s. The green horizontal lines represent the laser path at locations 1, 2 and 3. Note, the grey semicircle in the PIV images represents the `mask' used for image processing, not the cross-sectional area of the heating rod (which is shown in white in the LES images). From PIV results, an asymmetry in the rising plume is evident.}

  \label{simulation}
\end{figure*}

Fig. \ref{simulation} shows velocity magnitude and temperature contour plots obtained from PIV and LES in the plane of interest. There are several reasons to expect quantitative discrepancies between PIV and LES. For example, LES assumes a uniform heat flux distribution from the heating rod; however, we observed a marked non-uniformity in temperature employing an infrared camera in the lab (see Fig. \ref{fig:LES_IR}b). Nevertheless, a qualitative interpretation of these results is sufficient to substantiate our forthcoming arguments. 

Comparison of velocity magnitude obtained by LES and PIV shows a good qualitative agreement, particularly regarding the time of arrival of the plume's front to location 1. In the LES temperature plot, this hot water front is very clear (see Fig. \ref{simulation}a3), such that we can reliably associate it with the first peak detected via interferometry (see Fig. \ref{dynamic results}a). At this point, it is worth recalling that our method captures the average temperature change along the laser path in the water. Therefore, as the plume progresses upwards, the laser at location 1 travels through the thin central `neck' and lateral `arms' of the plume (Fig. \ref{simulation}b3), detecting a decrease of temperature, until the counter-rotating vortices at the edge of the arms reach this location (Fig. \ref{simulation}c3), leading to the detection of a second peak. To explain then the third peak in Fig. \ref{dynamic results}a (inset), we notice the asymmetry in the rising plume observed via PIV (Figs. \ref{simulation}b1 and \ref{simulation}c1), which is to be expected given the system's sensitivity to initial conditions \citep{narayan2017interferometric}, as well as imperfections, such as the heat flux non-uniformity discussed above. Because of this asymmetry, a third peak is detected, corresponding (in this case) to the left-arm eddy, which reaches location 1 with a time delay relative to the right-arm eddy (compare positions of dotted white circles in Figs. \ref{simulation}c1 and \ref{simulation}c2). Additional support for this interpretation comes from the fact that multiple repetitions of this experiment consistently revealed the presence of the aforementioned three peaks at locations 1 and 2 almost every time, but, crucially, not always. Sometimes only two peaks would be detected, which can be attributed to conditions when the plume's asymmetry was negligible. 
This method's sensitivity thus allows us to capture the signature of fine fluid dynamical patterns in a fully non-intrusive manner.

\subsection{Thermal gradient effect on interference visibility}

\begin{figure}[H]
  \centering
  \includegraphics[width= 0.95 \linewidth]{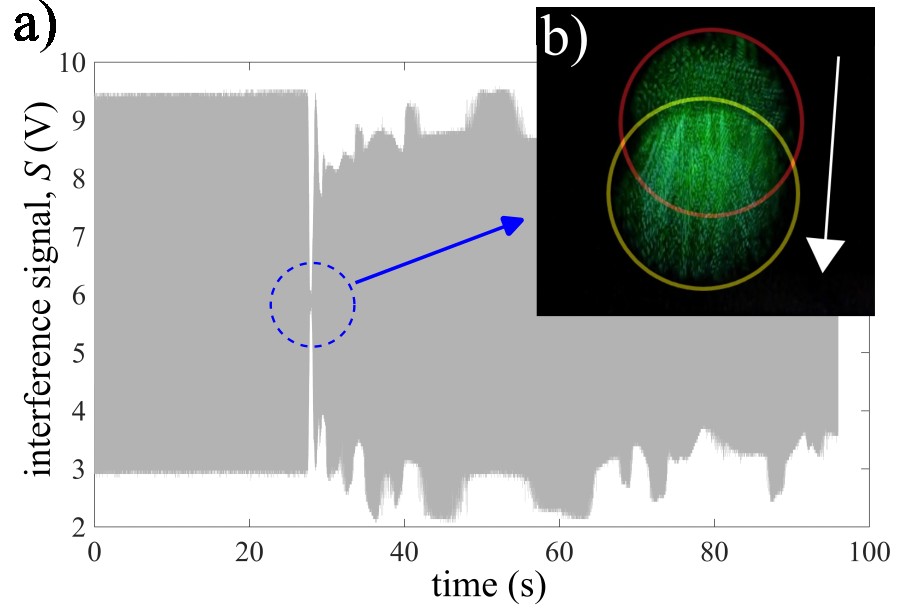}
  \caption{a) Interference signal detected by the PD at location 1, indicating the drop in visibility $\gamma$ (blue dashed circle) caused by the arrival of the plume's front (see inset in Fig. \ref{dynamic results}a). b) We use a camera to record the temporary change in the probe beam's position due to refraction, shown by the red (original/permanent position) and yellow (temporary/displaced position) circles; compare with inset in Fig. \ref{fig:simple_setup}c.}
  \label{beam}
\end{figure}

During the convective flow experiment, the interference visibility $\gamma$ (see Fig.~\ref{fig:simple_setup}b) decreases substantially when the thermal plume reaches the measurement location, as indicated by the blue circle in Fig.~\ref{beam}a. We claim that this effect is caused by the large vertical temperature gradient across the laser beam induced by the thermal plume's front, causing significant refraction of the probe beam and subsequent misalignment with the reference beam at the photodetector. This hypothesis is confirmed by recording the beam's position using a camera, the results of which are shown in Fig.~\ref{beam}b. Note that despite the reduced visibility periods observed in Fig.~\ref{beam}a, we are still able to track the same \change{visibility}{interference} peak and measure temperature change reliably in the convective flow setting. In other words, the visibility drop does not lead to accumulation of measurement errors, as discussed in Section \ref{sec:convec_vs_RTDs}. However, in other experimental settings (e.g. a stronger plume presenting a \add{much larger} temperature gradient\remove{closer to an actual thermal shock}), visibility reduction can lead to temporary data loss. 
This \add{in turn} makes visibility reduction due to thermal gradients a useful indicator of measurement performance. If the thermal gradient is excessive, our method \add{(unlike HI)} will provide no data, instead of yielding a distorted measurement. This feature distinguishes the present approach (SI) from holographic interferometry, where refraction (due to the thermal gradients in the fluid) can lead to phase reconstruction errors~\citep{Lira87} \add{(see also Section} \ref{secIntro}), and renders SI especially suitable for probing rapid temperature changes in three-dimensional flows like the one herein considered. 

\section{Discussion} \label{secDiscussion}
\add{We now provide a summary of the main strengths and limitations of the proposed technique, as well as an outlook of potential future applications and improvements.

While using relatively simple and economic instrumentation, standard interferometry has the potential to yield highly precise temperature measurements in a fully non-intrusive fashion. Moreover, the robustness of this technique, along with its small data processing requirements, render it attractive for real-time applications. 
However, the temperature measurements obtained represent the average along the beam path within the test fluid (the line of sight), which may limit application of this method to a set of specific problems (see below). What is more, the interferometer is very sensitive to ambient disturbances (such as mechanical vibrations) and, for optimal performance, may need to be realigned before the start of an experiment (which can be time-consuming and requires specific skills from the experimentalist).

Despite current limitations, the present technique could be employed in small-scale experiments where high precision and non-intrusiveness are desirable. For example, it may be used to characterise quasi-two-dimensional flows (where the dimension traversed by the probe beam is much smaller than the other two), yielding data that might potentially serve as benchmark for validation of Computational Fluid Dynamics models. Moreover, in 3D flows (like that in Section 4), several interferometers may be arranged so as to span a plane, potentially enabling the recovery of more localised temperature data. Due to the increase of precision with integration time (up to a certain point), this technique can also be helpful in long experiments with slow-evolving temperature where very high precision measurements (in the order of ${\rm\mu K}$) might be beneficial. 

In this paper, we have purposely applied this method to two problems with a significant disparity in nature and complexity (i.e. slow varying temperature in static water and a convective 3D flow), with the goal of better illustrating the strengths and weaknesses of our approach. However, our overarching aim is to promote within the community the adoption and adaptation of standard interferometry for the study of a wider range of problems in experimental fluid mechanics.}

\section{Conclusions} \label{secConclusions}

\add{We explore some of the merits and shortcomings of standard interferometry by applying it to two fluid problems: slow-varying temperature in static water and natural convective flow.} We detect \add{slow and} rapid \add{line-of-sight-average} temperature changes in water at mK sensitivity in a fully non-intrusive manner. This is \change{achieved}{enabled} by \change{employing standard interferometry, which detects}{the interferometer's high sensitivity to} minute temperature-induced changes in the fluid’s refractive index\remove{integrated over the probe beam path}.\remove{The merits of this technique are demonstrated both for slow-varying temperature in static fluid and for rapid temperature variations in natural convective flow.} We achieve \change{a remarkable accuracy}{a precision}  of about 1 mK per measurement with a sampling rate of 40 Hz. The high sensitivity and fast response time of our method \change{enable}{allow} us to capture the signature of fine fluid dynamical patterns that cannot be resolved with traditional methods such as RTD sensors. For example, in convective flow induced by a horizontal heating rod, we observe an asynchronous arrival of two counter-rotating vortices at the measurement location, revealing an asymmetric thermal plume, which we further corroborate by means of Particle Image Velocimetry and Large Eddy Simulations. 

Resolving \change{the}{potentially large} thermal \change{quasi-shock}{gradients} induced by \change{the}{a} rising plume is particularly challenging for any temperature measurement technique. For example, RTD sensors, in addition to being intrusive, are unable to capture this phenomenon due to their relatively slow response time, whereas Holographic Interferometry based methods cannot reliably operate in the presence of large thermal gradients\remove{inherent to the rising plume}. Conversely, we are able to \change{capture}{operate in the presence of} said \change{near-shock}{steep gradients} thanks to the fast response and robustness of our approach. 

\add{Despite its current limitations (mainly stemming from the line-of-sight-average nature of its measurements),} the technique herein described may be utilised to investigate \add{slow and} rapid fluid temperature variations in \change{a minuscule scale, such as energy dissipation in eddies, or for validating computational fluid dynamics (CFD) models, offering benefits over conventional methods due to its non-intrusive nature, small data processing requirements and the provision of real-time, highly accurate results.}{certain types of experiments (e.g. quasi-2D flows), generating high-quality data that might be used, for example, for validation of Computational Fluid Dynamics models. Moreover, as illustrated here, combination of this with other techniques (such as PIV), can enhance the analysis of a wide variety of problems in experimental fluid dynamics.}

\begin{appendices}

\section{PIV settings}
\label{piv appendix}
A green laser (532 nm) with an output power of 4.5 mW is used as the source, and a cylindrical lens is employed to generate the light sheet necessary for Particle Image Velocimetry. A high-speed CMOS camera (Baumer VCXU-15M) was utilised to capture particle images at 100 frames per second. A bandpass filter was placed in front of the camera to eliminate ambient light and ensure that only the scattered  green light from the tracer particles was captured. \add{Polyamide particles are used as tracer particles, with a diameter of around 55 $\mu$m and a density of 1.016 g/cm$^3$. Our field of view (FOV) is a square region that extends from the centerline of the heating rod to the uppermost water level, with a width equal to the inner width of the tank.} Preliminary tests indicated that the maximum velocity magnitude was approximately 10 mm/s. As suggested by \cite{lu2013high}, particle displacements between subsequent frames should not exceed 1/4 of the interrogation window size (8 pixels in our configuration) to reduce the number of pairing loss (loss of particle images within the interrogation window between subsequent frames). The time required for a particle to move 8 pixels is 0.07 s for the maximum expected velocity, which is much larger than the time between adjacent frames for our configuration (0.01 s), indicating that our chosen frame rate is suitable.

\begin{figure}[H]
  \centering
  \includegraphics[width= 1 \linewidth]{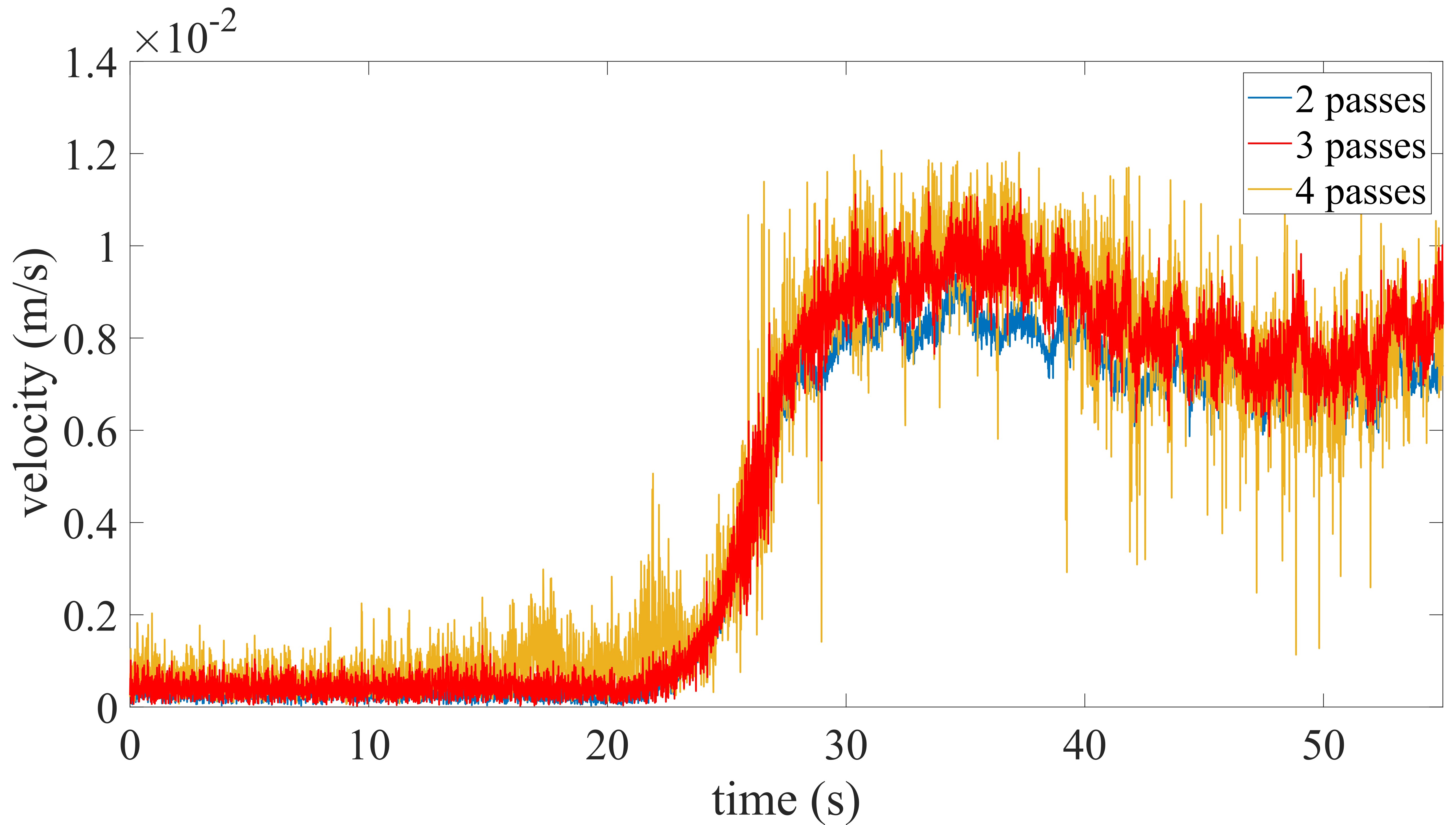}
  \caption{Velocity magnitude at a point along location 1 and just above the heater, obtained by using three different pass settings when processing PIV images.}
  \label{PIV}
\end{figure}

Image processing was performed using PIVlab \citep{thielicke2014pivlab}. A multi-pass interrogation algorithm with window deformation was implemented to improve the calculation accuracy of the velocity field. Three passes were used in total, which are 128$\times$128, 64$\times$64 and 32$\times$32 pixels with a 50\% overlap. To investigate the influence of the number of passes, we compare the velocity magnitude in a selected point (along location 1 and just above the heating rod) obtained by also using 2 passes only (128$\times$128 and 64$\times$64 pixels) and 4 passes (128$\times$128, 64$\times$64, 32$\times$32 and 16$\times$16 pixels). Results are shown in Fig. \ref{PIV}. We observe that 2-passes produces lower velocities compared to the other two settings, while 4-passes introduces the most noise. Thus, in this paper, we report results using the 3-pass interrogation algorithm. \add{The interrogation area size for each pass are 11.5 mm, 5.8 mm and 2.9 mm.}

\section{LES settings}
\label{les appendix}

The computational domain in the simulation is set to match our experimental settings, as detailed in Section \ref{set-up}. The  Navier-Stokes equations are solved using a Large Eddy Simulation (LES) approach with Boussinesq approximation, and the subgrid-scale (SGS) turbulence was modelled using the dynamic Smagorinsky model. The mesh grid has a total number of $40\times10^6$ hexagonal cells. The minimum grid size is 0.03 mm, and the non-dimensional wall distance $y^+$ on the heating rod is 0.023. For the boundary conditions, we assign no-slip conditions to the side and bottom walls, while the top wall is designated as a free-slip boundary, considering the fact that any free surface level fluctuations that may occur will be negligible in magnitude compared with the water depth \citep{mcsherry2017large}. We set the heater's boundary condition as a no-slip wall and define a constant heat flux of 1504 Wm$^{-2}$, in line with the experiment's power output of 12 W. The initial temperature is set to 22$^{\circ}$C everywhere. Simulations are carried out in ANSYS Fluent 21.2. The numerical algorithm and solver settings were adapted from \cite{ma2021large}, who also analysed a similar case study using LES. A pressure-based solver was employed, and SIMPLEC scheme was used for the pressure-velocity coupling scheme. Discretisation of the gradients and pressure equation was conducted by least squares cell based and body force weighted scheme, respectively. For the momentum and energy equations, boundary central differencing and second order upwind schemes are adopted, respectively. Time marching is achieved by a bounded second order implicit scheme. The simulations were run on the University of Southampton's Iridis 5 supercomputer.

\begin{figure}[H]
  \centering
  \includegraphics[width= 1 \linewidth]{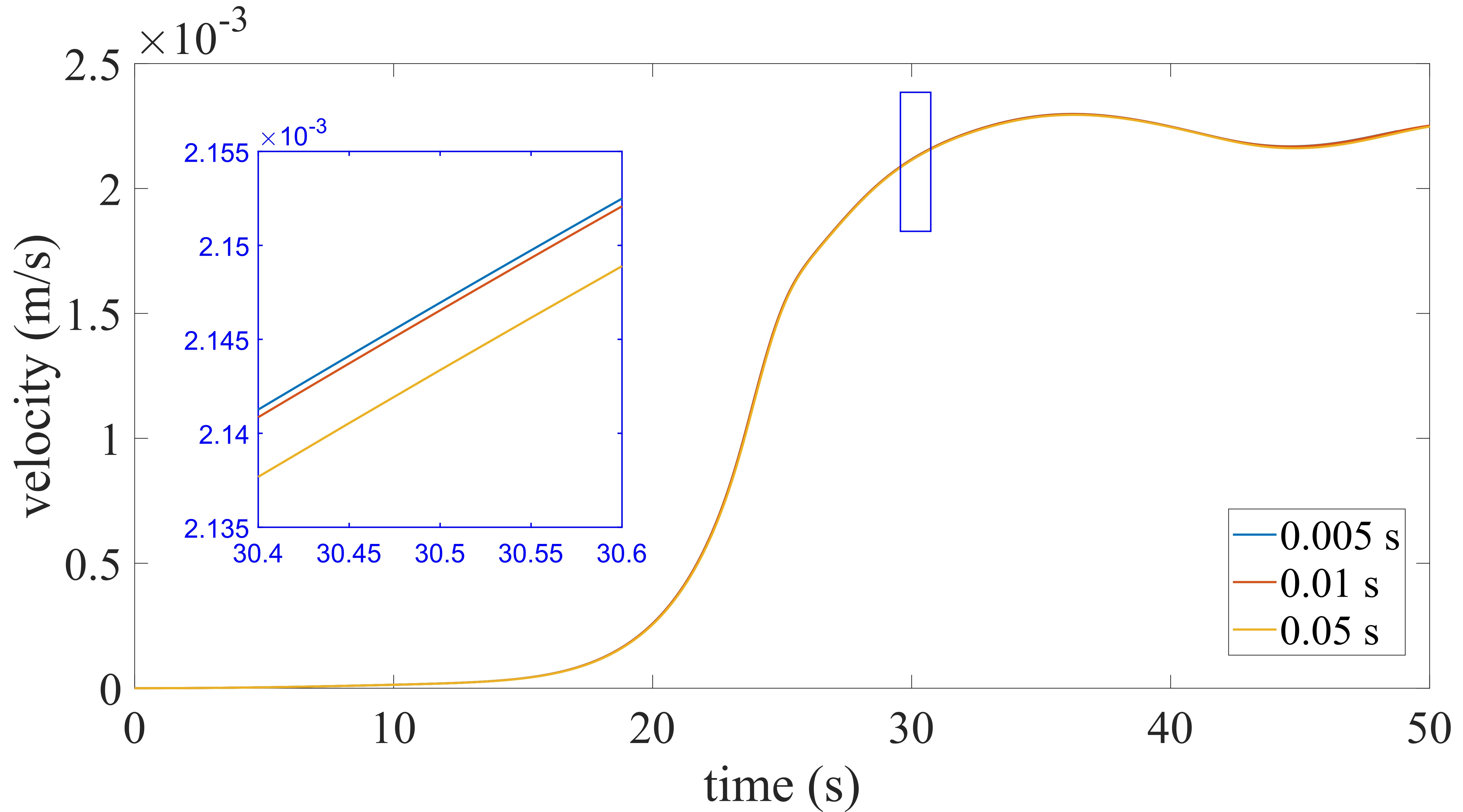}
  \caption{LES-obtained mean velocity magnitude along location \change{2}{1} (i.e. averaged over the beam path at location \change{2}{1}) for three different time steps.}
  \label{time step}
\end{figure}

To determine the integration time step, Fig. \ref{time step} illustrates the mean velocity magnitude along location \change{2}{1} (that is, averaged over the beam path at location \change{2}{1}) for three different values of the time step. The decrease in time step from 0.01 to 0.005 s yields a marginal mean difference of only 0.24\%. In contrast, a reduction in the time step from 0.05 to 0.01 s results in a larger mean difference of 1.79\%. Therefore, we conclude that the benefits of the additional computational time and resources required for smaller time steps beyond 0.01 s are not justified. Hence, we determine that a time step of 0.01 s provides sufficient precision for our needs and was consequently selected for our simulations.

\begin{figure}[H]
  \centering
  \includegraphics[width= 0.55 \linewidth]{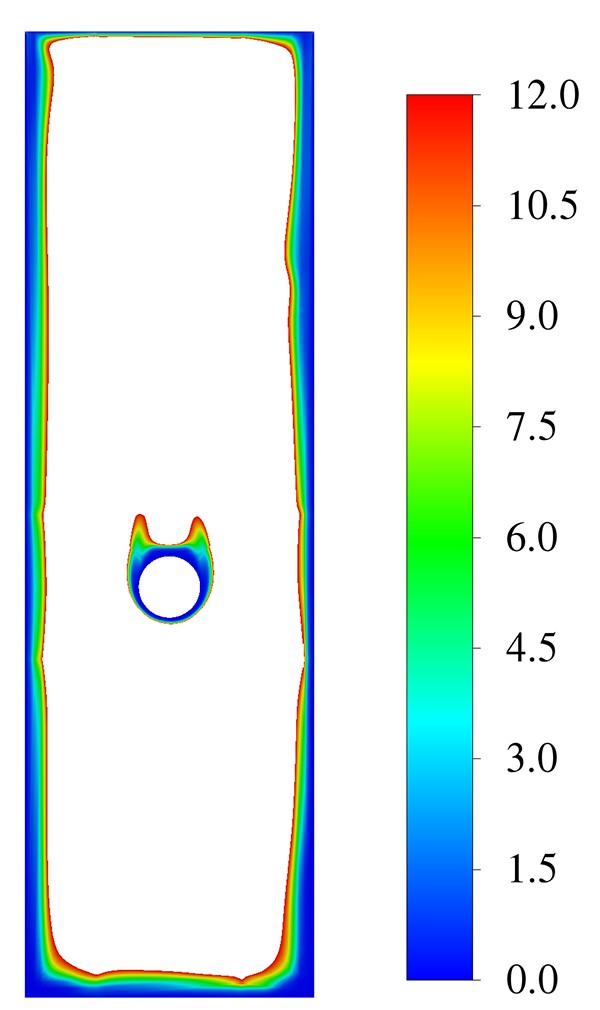}
  \caption{Contour of the ratio between the RANS integral length scale $l_1$ and the filter width $\Delta$ at 40 s after start of the experiment; the white area complies with the $l_1/ \Delta >12$ rule. \add{Same cross-section as in Fig.} \ref{simulation}, \add{but spanning the whole height of the tank}.}
  \label{grid size}
\end{figure}

The quality of the LES computational grid was judged based on a Reynolds-Averaged Navier-Stokes (RANS) simulation of the same case. The turbulent integral length scale, $l_1=k^{3/2} / \varepsilon$ (where $k$ and $\varepsilon$ are the turbulent kinetic energy and energy dissipation rate, respectively), was derived from the RANS simulation and served as a benchmark for assessing the quality of the mesh. The LES results are deemed satisfactory if more than 80\% of the turbulent kinetic energy is resolved. This goal can be achieved by ensuring the filter width, $\Delta$, is smaller than $l_1/12$, as suggested by  \cite{pope2000turbulent}, where $\Delta$ is computed according to the volume of the computational cell, $V$, using $\Delta=V^{1/3}$. Figure \ref{grid size} shows the ratio $l_1/\Delta$ computed for the grid used in our LES model. The contour range is confined between 0 and 12, such that the grid size in the white area complies with the $l_1/ \Delta >12$ rule, and therefore, the core region of the domain possesses an acceptable resolution. The non-dimensional wall distance, or $y^+$, on the heater and the walls is 0.023, which is sufficiently low to ensure a well-resolved LES near the wall.

\end{appendices}

\section*{Declarations}

\begin{flushleft}
\textbf{Ethical Approval} not applicable
\end{flushleft}

\bigskip

\begin{flushleft}
\textbf{Competing interests} The authors have no competing interests to declare that are relevant to the content of this article.
\end{flushleft}

\bigskip

\begin{flushleft}
\textbf{Authors' contributions} SM and JAZ conceived the research idea. XG performed the experimental work and LES, and processed all data. All authors discussed the results and contributed to the final manuscript.
\end{flushleft}

\bigskip

\begin{flushleft}
\textbf{Funding} Partial funding was received from the University of Southampton. XG was partly funded by the University of Southampton through a Presidential Scholarship.
\end{flushleft}

\bigskip

\begin{flushleft}
\textbf{Availability of data and materials} Data available from the authors upon request
\end{flushleft}

\bigskip

\begin{flushleft}
\textbf{Acknowledgements} The authors acknowledge the use of the IRIDIS High Performance Computing Facility, and associated support services at the University of Southampton, in the completion of this work. \add{The authors also thank two anonymous reviewers, whose thorough feedback has helped improve the overall quality of this paper.}
\end{flushleft}

\bigskip

\bibliography{sn-bibliography}

\end{document}